\documentstyle[epsf]{elsart}

\newcommand{\be}{\begin{equation}}
\newcommand{\bef}{\begin{figure}}
\newcommand{\eef}{\end{figure}}
\newcommand{\hmp}{h^{-1}Mpc}

\newcommand{\ee}{\end{equation}}
\def\spose#1{\hbox to 0pt{#1\hss}}
\def\ltapprox{\mathrel{\spose{\lower
3pt\hbox{$\mathchar"218$}}
 \raise 2.0pt\hbox{$\mathchar"13C$}}}
\def\gtapprox{\mathrel{\spose{\lower
3pt\hbox{$\mathchar"218$}}
 \raise 2.0pt\hbox{$\mathchar"13E$}}}
\def\inapprox{\mathrel{\spose{\lower
3pt\hbox{$\mathchar"218$}}
 \raise 2.0pt\hbox{$\mathchar"232$}}}

\begin{document}

\begin{frontmatter}
\title{CORRELATION PROPERTIES OF 
CLUSTER DISTRIBUTION }
\author[Roma,Infm,Calabria]{M. Montuori} 
\author[Roma,Infm]{F. Sylos Labini},
 \author[Roma,Infm]{A. Amici}
\address[Roma]{Dipartimento di Fisica, Universit\`a di Roma
``La Sapienza'' P.le A. Moro 2, I-00185 Roma, Italy.}
\address[Infm]{INFM, Sezione di Roma 1}
\address[Calabria]{Dipartimento di Fisica, Universit\`a
 della Calabria, Italy}

\begin{abstract}
We analyze subsamples of Abell and ACO cluster catalogs, in order 
to study the spatial properties of the large scale matter distribution.
The subsamples analyzed  
are estimated to be nearly complete and are the 
standard ones used in the correlation analysis by various authors.
The statistical analysis of cluster 
correlations is performed without 
any assumptions on the nature of the distribution itself.
The cluster samples show fractal correlations 
up to sample limits ($\approx 70 h^{-1} Mpc$) with fractal 
dimension $D \approx 2$, without any tendency towards 
homogenization.
Our analysis shows that the standard correlation 
methods are incorrect, since they give 
a finite correlation length for a distribution that does not possess one.
In particular, the so called correlation length $r_{0}$ is shown to be simply 
a fraction of the sample size.
 Moreover we conclude that galaxies and clusters are two 
different representations of the same self similar structure and 
that the correlations of clusters are the continuation of 
those of galaxies to larger scales.  
\end{abstract}
\end{frontmatter}
 
\section{Introduction}

The determination of the matter distribution
in space represents a crucial test for the initial 
hypothesis of the main cosmological theories
\cite{ba94}.
Both standard Friedmann and Steady State models 
assume, beyond a certain scale $\lambda_{0}$, 
the homogeneity of the matter distribution.
This assumption can be tested, properly investigating 
the spatial distribution of galaxies and clusters. 

With respect to the galaxy catalogs, cluster
surveys offer the possibility to study the large 
scale structure of the matter distribution in volumes 
much larger, reaching depths 
beyond $z\approx 0.2$
 and extending all over the sky.
Using clusters, one can trace matter distribution 
with a lower number of objects with respect to 
the galaxies 
in the same volume.
For example, in the northern hemisphere we know 
$\approx 10^{6}$ galaxies up to $z\approx 0.15$ 
that correspond to $\approx 500$ rich clusters.
However the problem of cluster catalogs
is their incompleteness, since clusters
are identified 
as density enhancements in galaxy angular surveys and 
their distance is usually determined through the 
redshift measurement of one or two galaxy member.
It is clear in fact that only the measurements of every 
clusters (and member galaxies) would allow to construct 
truly volume limited samples.
Moreover, as we show in the following,
there is a strong arbitrariness in the 
definition of a {\it cluster}: such an arbitrariness
can be avoid using the methods of modern statistical mechanics.

Cluster distribution 
shows large inhomogeneities and huge voids.
Tully 
(\cite{tu86}, 
\cite{tu92}),
 investigating 
the spatial distribution of  Abell and ACO 
clusters up to $300 h^{-1} Mpc$, 
observes the presence of structures 
on a scale of $0.1c$, lying in 
the plane of the Local Supercluster.
Galaxies clump in clusters and clusters in superclusters
with extensions comparable to the largest 
scales of current samples.
Catalogs of superclusters,
 compiled by various authors  
(\cite{ei94},   
\cite {bah83}, 
\cite {bb85},
\cite {po92},
\cite {zuc93})
show the presence of correlated structures 
with extension ranging from few $ Mpc$ 
to $150  h^{-1} Mpc$.
Complementary to the presence 
of large structures is the existence of 
voids. Several authors 
(\cite {bah83},
\cite {bb85}, 
\cite {kf91}, 
\cite {ha94}, 
\cite{ei94}) have 
investigated the shape and the dimension of voids 
determined by rich clusters and 
superclusters. The dimension of voids is 
usually defined 
as diameter of empty sphere containing 
no clusters.
As for clusters, the 
definition of voids implies 
a certain degree of 
arbitrariness, since 
one can consider only 
spherical voids or ellipsoidal voids and so on. 
In particular Einasto et al. 
(\cite {ei94}) 
find that in their 
cluster sample (up to $ z\approx 0.1$), the median radius of 
 voids is $50 h^{-1} Mpc$. However, the observed 
voids have elongated 
shapes and in one dimension they can exceed this value.
 The authors observe in fact a giant void of 
$300 h^{-1} Mpc$ of length.
A more detailed study 
on a single void is performed by 
\cite{li95}, which have investigated the 
distribution of galaxies in and around the 
closest one, the Northern 
Local Void. 
The authors found that the dimension of the 
voids depends on the objects which have been 
used to defined them: voids defined 
by clusters are larger than galaxy defined voids, which 
are larger than faint galaxy defined voids.
There is, then, a hierarchy of voids.
However voids seem to be 
real and not 
due to observational incompleteness. 
Search of extremely faint dwarf 
galaxies in voids has given, up to now, 
negative results (
\cite{hop94}, see refs in 
\cite {li95}).
Moreover one observes an increase of 
the void size with sample size, that 
may correspond to a 
a self similarity in the 
distribution of voids.
Even in this case we show that the modern methods 
of statistical mechanics allow a study of void 
distribution does not depend on 
any assumption of arbitrariness.

The observation of large scale structures like 
superclusters and voids, extending up to the 
limits of the samples, clearly 
make questionable  the existence of 
an average density: the average 
density may be not a well defined intrinsic property  
of the system and the assumption of homogeneity 
 for the distribution must be checked and not assumed.
Usually, the presence of inhomogeneities 
on the scale of sample analyzed is interpreted 
as incompleteness of the samples themselves.
However, up to now, 
 better and more 
extensive observations 
have confirmed the reality 
of these structures, 
making the observed agglomeration 
sharper and not filling the voids.

An argument usually proposed 
to reconcile the observed inhomogeneities 
to the supposed large scale homogeneity, 
is that the structures 
are indeed real, but their amplitudes are small; 
in other words 
the density contrast $\delta N/N$ ,i.e. the 
ratio of the density fluctuations with respect 
the average density, tends to zero at the limits of the 
sample (
\cite{ps91}, 
\cite {pee93},
\cite {ps94}).
 One expects, therefore, that going a bit further, 
 the homogeneity would be  
eventually reached. A part from the 
fact that this expectation has been 
systematically disproved, the argument is 
{\it conceptually wrong}.
As reported in 
\cite {ba94}, $\delta N/ N $ tends to 
zero at the limits of the sample 
for any system and can not be interpreted as tendency 
toward homogenization.
This is because the average is computed within the sample 
volume.

The standard measure of the 
irregularities in the space distribution is the 
dimensionless autocorrelation function 
(CF) $\xi(r)$.
This is the density-density correlation 
normalized by the average density of the 
system.
If the average density is not an 
intrinsic property of the system,
unique and independent, 
the results are spurious 
(\cite {pie87},
\cite {co88}, 
\cite {co92}).
The basic point is that the 
use of the standard CF assumes {\it implicitly} that 
the homogeneity is well reached within the sample size, 
and consequently it can not test whether it happens or not.

For both galaxy and cluster 
distributions,
$\xi(r)$ function is
found to be a power law: 
\be
\xi(r) \approx (r_{0}/r)^{\gamma}
\ee
both the exponent and the amplitude $r_{0}$
are uncertain. The exponent $\gamma$ 
generally appears consistent with the value 
$1.8$ observed for galaxy autocorrelation function
(
\cite {kir78},
\cite {pee80},
\cite {da83},
\cite {sha83}, 
\cite {dep88}).
The quantity $r_{0}$ is the so called 
"correlation length" and is 
defined as $\xi (r_{0})=1$.
 
In general high values for $r_{0}$ 
($20 - 25 h^{-1} Mpc$) 
are obtained 
for samples that contains more rich 
Abell/ACO clusters
(
\cite{ab58}, 
\cite{bah83},
\cite{kl83},
\cite{bah88},
\cite {ab89},
\cite{hu90}, 
\cite{po92},
\cite{ca92}),
although there is a 
debate whether these high 
values might be produced by 
systematic biases present
in the Abell/ACO catalog.
The proposed corrections for 
these effects produce a lower 
$r_{0}$ of $ \approx 14 h^{-1} Mpc$
(
\cite{su88},
\cite {dek89}, 
\cite {su91},
\cite{ef92},
\cite{va96}).
 
In conclusion, for the 
Abell/ACO samples analyzed so far:
\be 
14 h^{-1} Mpc \ltapprox r_{0} \ltapprox 25 h^{-1} Mpc
\ee
 
According to the standard analysis, $r_{0}$ should be 
the correlation length for the distribution 
and its physical meaning is the following: 
below this value, 
clusters should be strongly 
correlated, while correlations should 
became small and eventually negligible at 
scale of the order of twice 
this length.
At this distance the distribution should became 
rather smooth and homogeneous.
The observational evidences go instead exactly 
in the opposite direction, since  
larger samples show larger correlated structures.
The result of the standard 
statistical analysis, i.e. 
a little correlation length, 
seems in contradiction to 
the observational evidences. 
Moreover clusters are found to be 
more clustered than galaxies, 
 for which  the standard 
analysis obtains a correlation 
length $r_{0}$ $\approx
5 h^{-1} Mpc$
(
\cite {gro77}, 
\cite{pee93}).
This mismatch between galaxy and cluster correlations is 
a puzzling feature of the usual analysis; clusters, 
in fact, are made by galaxies and  
many of these are included in the galaxy catalogs 
for which  $r_{0} \approx 5 h^{-1}\: Mpc $
 is derived.
It is therefore necessary to assume that cluster galaxies 
have fundamental differences with respect these that are 
not part of a cluster. This concept has given rise to the so called
richness clustering relation 
(\cite{bah83},
\cite{bah86},
\cite{bah92},
\cite{bah94});
according to this, 
objects with different mass or morphology 
would segregate from each other and produce 
different correlation properties; 
in other words, the correlation length 
of a given class of objects $i$ is 
$r_{0i} \approx 0.4<n>^{-1/3}_{i}$,
where $<n>_{i}$ is relative mean spatial density.
This hypothesis has been applied to explain the increase
of $r_{0}$ going from APM and $X-ray$ less rich
clusters to the Abell more rich clusters. 
Szalay and Schramm 
\cite{sz85}
 have 
interpreted the scaling of  $r_{0}$  as signature of 
scale invariance properties of the 
distribution. However, as pointed out by 
\cite {co92}
  (hereafter CP92), 
 this is not correct since 
for a self-similar distribution $r_0$ is proportional
to $R_s$, where $R_{s}$ is the {\it effective} sample 
radius (see sec.5) and not to 
$ <n> ^{-1 /3}$. In effect, statistical analysis
of sparser objects requires larger
sampling volumes.
As the following analysis will show, 
cluster distribution
is fractal up to the sample size; then we can
reinterpret the scaling of $r_0$ as
an increase of $R_s$.
 
The criticisms of the standard analysis have been raised by 
Pietronero and collaborators (
\cite {pie87}, \cite {co88},
CP92).
In CP92, the authors analyze, with methods of modern 
statistical mechanics, 
the CfA1 catalog of galaxies 
(
\cite {hu83}) and a sample of clusters from 
Abell catalog 
(
\cite {ab58}).
This analysis, differently from the $\xi(r)$, has no 
a priori assumptions on the nature of 
the distribution and is commonly used 
in modern statistical mechanics,
where one has to face non-analytical 
distributions. 
The results 
are that the galaxy and cluster
distributions are not homogeneous 
within the sample limits, but 
on the
contrary, fractal up to 
the sample limits for 
statistical analysis: these 
are  
$ \approx 20 h^{-1} Mpc $
for galaxy and to $ \approx 60-80 h^{-1} Mpc$
for cluster in the samples analyzed.
The analysis is performed to greater distances 
for clusters simply because the volume 
of cluster sample is larger than 
the galaxy one.
In other words, there is not 
any scale of homogeneity and 
 correlation 
properties of clusters are just the continuation of the 
galaxy correlations to larger scales. The mismatch
between the cluster and galaxy correlation is 
due to a mathematical inconsistency 
(the use of $\xi(r)$) and it is 
automatically eliminated 
using the correct statistical analysis.
 
Recently such an analysis has been performed also on 
several other  
galaxy surveys, namely Perseus Pisces redshift
catalog, LEDA, ESP: 
these results confirm and extend 
those of CP92.
Galaxy 
distribution in Perseus Pisces 
catalog is not homogeneous, 
but fractal 
with the fractal dimension 
$D \approx 2$ up to 
$30  h^{-1} Mpc $ (
\cite {sym96}, \cite {syl96}, \cite {syp96}, 
\cite {syloam96}); 
in LEDA database, fractal  up to
$150 h^{-1} Mpc$ with dimension 
$D \approx 2 $ (
\cite {din96};,
\cite {am96}).
ESP catalog requires a slight 
different analysis, which implies 
a lower degree of statistical validity.
With these remarks, ESP catalog 
shows a fractal behaviour with 
the same dimension ($D \approx 2$)
 up 
$800-900 h^{-1} Mpc$
(
\cite {ba94}).
Putting all these results together, one 
obtains power law (i.e. {\it fractal}) correlations 
for galaxy distribution, 
extending from few $ Mpc$
up to $800-900 h^{-1} Mpc$
\cite {sym97}.
Consequence of fractal correlations is
that $r_{0}$, derived from $\xi(r)$, 
is not the correlation length,
but simply a fraction of the effective sample radius $R_{s}$.
According to this, $r_{0}$ scales in 
all samples  
with  $R_{s}$.

In these paper we present results of the statistical analysis 
on various samples of Abell/ACO clusters.
 In section 2 we give a description of the 
samples analysed.
In section 3 we discuss the samples completeness 
and the behaviour of the density versus radial distance.
In section 4 we briefly introduce the 
basic properties of fractal distributions.
In section 5 we present the results of the 
correlation function $\Gamma(r)$ analysis.
In section 6 we study the standard $\xi(r)$
function and we analyse the 
sample depth dependence of the 
correlation length $r_{0}$ and finally 
we summarize 
the results and draw our conclusions.

\section{Description of the data and subsamples}

The study of the spatial distribution
of clusters requires a complete sample of 
clusters over a large volume.   
There are many available extensive
catalogs of clusters from which one may try
to extract complete subsamples. 

Here we have analyzed subsample 
extracted from Abell catalog and ACO catalog.
From Abell catalog we have analyzed 
the
Postman sample (
\cite{po92}); 
this is a sample of 351 clusters with 
the tenth ranked galaxy magnitude ($m_{10}$)$\le 16.5$.
The sample includes 
all such clusters which lie north of 
$\delta=-27^{\circ} 30'$. The typical 
redshift of a cluster with $m_{10}\le 16.5$
is $z\approx 0.09$. To this redshift, incompleteness
in the Abell catalog is considered insignificant 
(
\cite{po92}).
About half of clusters have 
the redshifts based on at least 
three independent galaxy spectra, 
$25 \%$  
have redshift determined 
from two galaxy spectra
and the last 
$25 \% $  
from a single spectrum (
\cite {po92}).

Postman et al. (1992)
\cite{po92} 
have defined a so called 
statistical subsample, which consist of 208 
clusters with $z\le 0.08$ and $|b|\ge 30^{\circ}$.
This subsample has been shown to be minimally 
affected by selection biases.

Bahcall \& Soneira 1983 sample (
\cite {bah83}, BS83 hereafter): this sample 
includes 104 clusters of distance 
class $D\le4$ ($z\ltapprox 1$), richness class
$R\ge 1$ and located at high 
galactic latitude ($|b|\ge 30^{\circ}$)

From ACO catalog
\cite{ab89}
 we have selected all the cluster
with measured redshift   
in according to the following constraints:
$m_{10} \le 16.4,  b \le - 20^{ \circ}, \delta \le -17^{\circ}$
(
\cite {pli91}, \cite {pli92}). 
The sample contains 139 clusters up to 
the limiting distance of $\approx 930 h^{-1} Mpc$.

\section{Sample completeness}

The completeness of the samples 
is usually
estimated from the 
density of the clusters as function of the 
redshift. 
In Fig.1, it is shown the density 
in the samples analysed.
The density $n(r)$ shows large fluctuations 
followed by 
a power law decay as  $ \sim r^{-3}$.
Computation of the density in bins produce 
larger fluctuations, because this is 
a differential 
quantity more noisy than the integral  density 
 $n(r)$.
Usually the 
fluctuating behaviour of the density up to 
$z\ltapprox0.1$ is interpreted as 
a flat one, i.e. the density of the sample is 
considered to be constant and the distribution 
homogeneous; consequently the sample 
is considered complete and homogeneous 
up to this distance and incomplete 
beyond. 
The decay $r^{-3}$ is, in fact, due 
incompleteness of the samples, since 
the number of clusters is nearly 
constant, but 
the volume sampled increases.
We note that up to the 
beginning of the incompleteness 
region, all the samples contain few clusters; 
the sparsest is the BS83 sample: the 
northern galactic part contains 53 
clusters up to $230 h^{-1} Mpc$ 
and the 
southern part 24.
The Postman sample 
\cite{po92} 
contains 120 clusters up to 
$z\approx 0.08$ in the northern part 
and 88 in the southern one.
The ACO sample 91 up to the $150 h^{-1} Mpc$.
The basic point is that $n(r)$ is not an averaged 
quantity, i.e. is the density computed 
from the vertex.
In this situation 
at small scales, there are no clusters and 
the density is zero; going a little bit further, 
one starts to see some clusters, and the density 
shows large fluctuations because the statistics is 
low. 
At larger distance, the fluctuations 
reduce for the increasing of the 
statistics, but the average 
behaviour is apparent only sampling a 
quite large range of scale. 
In a homogeneous 
distribution, the amplitude 
of fluctuations 
reduce with scale,  
while in a fractal these 
have the same amplitude in a log scale,
because the structure is self-similar and 
is made by fluctuations (see sec.4).
Hence, to recover the mean behaviour, one 
has to sample a quite large range of scale.
   
\bef
\epsfysize 10cm
\centerline{\epsfbox{fig1n.post}}
\caption{
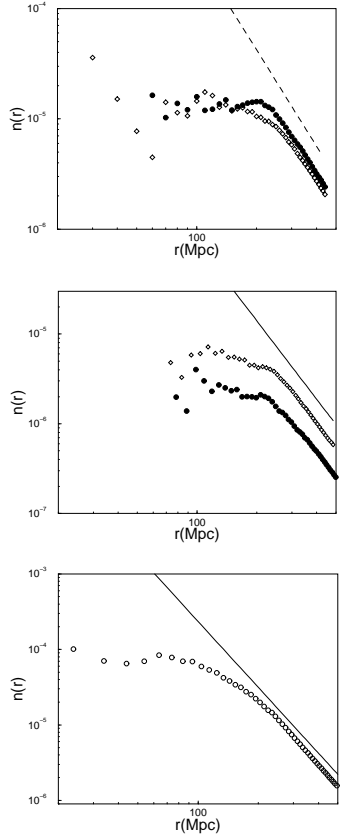
The radial density $n(r)$ 
from the vertex for 
a) Abell catalog Postman sample,
north (diamonds) and south (filled circles) 
b) Abell catalog BS83 sample north (diamonds)  and south (filled 
circles)
c) ACO catalog sample (empty circles).}
\eef
In Fig.1 we see that the 
 range of scale, over which 
we can measure $n(r)$,  
is quite small ($\approx 50 \hmp$) and 
in our opinion is not 
enough to recover the 
average behaviour of 
the density.

The study the properties of cluster spatial distribution 
requires an almost complete sample; 
then it is necessary to exclude or 
correct the 
incompleteness region.
One way to obtain this in the standard approach is to 
 assume an homogeneous distribution of clusters 
up to the sample limits. The observed distribution in the 
redshift space 
is then weighted with a selection function $p(z)$, that 
is the ratio between the observed counts of clusters 
in the volume $dV(z)$ at redshift $z$ and those 
expected from an homogeneous distribution.
In our analysis, we want to avoid any assumption on the distribution itself 
and for this reason 
we will limit our analysis up to a depth corresponding 
to the beginning of the incompleteness region, 
without correcting by means of any selection function.

Another selection effect must be taken in account, regarding 
to the observed depletion of the surface mean density 
of clusters at low galactic latitude (to $|b| \ltapprox 30^{\circ}$). 
This is probably due to 
obscuration and confusion with high-density regions of stars 
of our galaxy (
\cite {bah88}).
As in the case of redshift incompleteness,
one way to overcome this 
incompleteness is to weigh the observed distribution with 
a latitude selection function 
$P(b)$, that is the ratio between
observed surface cluster density at latitude $b$ and the 
expected from an homogeneous one. 
The normalization of this selection function 
is arbitrary, 
because it depends from the real density.
Another way is to limit the sample to high galactic latitude region.
We will adopt this standard procedure
(i.e. $|b|\ge 30^{\circ}$ for Abell catalog and $b < -20^{\circ}$
for ACO), that has no assumptions 
with only the inconvenient to slightly limit 
the sample. 
 
For the Postman sample \cite {po92} 
the distance of completeness 
is estimated to be roughly 
$ 230 h^{-1} Mpc$
corresponding to $z\le 0.08$,
for the BS83 sample the same 
distance, while for ACO sample
is $\approx 150 h^{-1} Mpc$. 
All these limits have been estimated from the 
behaviour of the density versus the distance
(Fig. 1).

\section{Properties of fractal distributions}

In this section we mention the essential properties 
of fractal structures because they will be necessary for
the correct interpretation of the statistical analysis. 
However in no way these properties are assumed
 or used in the analysis itself.
A fractal consists of a system in which more 
and more structures appear at smaller and 
smaller scales and the structures at small 
scales are similar to the ones at large scales.
The self similarity of these structures is 
then incompatible with analyticity.
Standard mathematical tools based on 
 analytical functions can not 
characterize these distributions.
The first quantitative description of these forms is 
the metric dimension.
One way to determine it, is the mass-length
method. Starting from an 
 point occupied by an object, we count how 
many objects $N$
("mass") are present within a volume 
of linear size  $r$ ("length") (
\cite{man82}):
\be
\label{l1}
N(r) = B\cdot r^{D}
\ee
 $D$ is the fractal dimension 
and characterizes in a quantitative way
 how the system fills the space.
The prefactor $\:B$ 
depends to the lower cut-offs of the distribution; these
are related to the smallest scale above 
which the system is self-similar 
and below which the self similarity 
is no more satisfied.
In general we can write:
\be
\label{l6}
B = \frac {N_{*}} {{r_{*}}^{D}}
\ee
where $r_{*}$ is this smallest scale 
and $N_{*}$ is the number of object
up to $r_{*}$.
For a deterministic fractal this relation is exact, while 
for a stochastic one  it is satisfied in an average sense.
  
Eq.(\ref{l1}) corresponds to a smooth convolution 
of real $N(r)$, that is a very fluctuating 
function; a fractal is, in fact, characterized by  large
 fluctuations and clustering at all scales.
We stress that eq. (\ref{l1}) is valid 
in general; for an homogeneous distribution, for example, 
one has $D=3$. 

From eq.(\ref{l1}), we can compute the 
average density $\:<n>$ for a sample of
 radius $\:R_{s}$ which contains a portion 
of the structure with dimension $D$. The sample volume 
is assumed to be a sphere ($\:V(R_{s}) = (4/3)\pi R_{s}^{3}$) and therefore
\be
\label{l2}
<n> =\frac{ N(R_{s})}{V(R_{s})} = \frac{3}{4\pi } B R_{s}^{-(3-D)}
\ee
If the distribution is homogeneous, $D = 3$ and the
average density is constant and independent from the sample 
volume; for a fractal, the average density 
 depends explicitly on the sample 
size $\:R_{s}$ and it is not a meaningful 
quantity.
 It is a decreasing function of the sample size and 
$<n> \rightarrow 0$ for $ R_{s} \rightarrow \infty$.
It is important to note that eq. (\ref{l1}) holds from every point of the 
system, when considered as the origin.
This feature is related to the non-analyticity of the 
distribution.
In a fractal distribution every observer is 
equivalent to any other one, i.e. it holds the property 
of local isotropy around any observer (
Sylos Labini 1994
\cite {sylo96}).
We can define the conditional density from an 
occupied point $i$ as:
\be
\label{l3}
\Gamma (r)_ {i}= S(r)^{-1}\frac{ dN(r)}{dr} = \frac{D}{4\pi } B r ^{-(3-D)}
\ee
where $\:S(r)$ is the area of  a spherical shell of radius $\:r$.
$\Gamma(r) _ {i}$ is then the density at distance $r$ from the 
$i-th$ point  in a shell of 
thickness $dr$.
As for eq.(\ref{l1}), eq.(\ref{l3}) corresponds 
to a smooth convolution of fluctuating quantity, i.e. 
the conditional density from one element of the distribution.
Usually the exponent that defines the decay
 of the conditional density $(3-D)$ is called 
the codimension and it corresponds to the 
exponent $\:\gamma$ of the galaxy distribution.

\section{The conditional average density}

The correlation function suitable to study 
homogeneous and inhomogeneous 
distribution 
is described by (
\cite {pie87}; CP92)
\be
\label{g1}
G(r) = <n(\vec{r}+\vec{r}_{i})n(\vec{r}_{i})>_{i} \approx r^{-\gamma}
\ee
where the exponent $\:\gamma = 3-D$
(in 3-dimensional space)(Eq. \ref{l3}) and 
the index $i$ of the average means that 
this is performed on all the 
occupied points $r_{i}$ 
of the system. 
If the sample is homogeneous,$D=3$, 
 $\:G(r) \approx <n>^2$ and
then is constant; if the sample has correlations on 
all scales, it is fractal, $D < 3$, $\:\gamma > 0$ and 
$\:G(r)$ is a {\it power law}.
For a more complete discussion we refer the reader to 
CP92. We can normalize the $G(r)$ to the size  of the 
sample under analysis and define, following CP92:
\be
\label{g2}
\Gamma(r) = \frac{<n(\vec{r}+\vec{r}_{i})n(\vec{r}_{i})>_{i}}{<n>} 
= \frac{G(r)}{<n>}
\ee
where $\:<n>$ is the average density of the sample.  This normalization
 does not introduce any bias even if the average
density is sample-depth dependent, 
as in the case of fractal distributions (Eq.\ref{l2}), 
because it represents 
only an overall normalizing factor. 
The $\Gamma(r)$ (Eq. \ref{g2})
can be computed by the following expression
\begin{eqnarray}
\label{g3}
\Gamma(r)& =& \frac{1}{N} \sum_{i=1}^{N} \frac{1}{4 \pi r^2 \Delta r} 
\int_{r}^{r+\Delta r} n(\vec{r}_i+\vec{r'})d\vec{r'}=\nonumber\\
& &= \frac{1}{N} \sum_{i=1}^{N} \Gamma(r)_ { i} =
\frac{D}{4 \pi} B r^{3-D}
\end{eqnarray}
where $N$ is the number of objects in the sample.
This is called the {\it conditional average density} (CP92).
$\Gamma(r)$ is the average of $\Gamma(r)_{i}$ and 
hence it is a smooth function away from the 
lower and upper cutoffs of the distribution ($r_{*}$ and 
the dimension of the sample).
From eq.(\ref{g3}) we see that 
$\Gamma(r)$ is independent from the 
sample size, depending only by 
the intrinsic quantities of the distribution 
($B$ and $D$). 
It is also very useful to use the average density
\be
\label{g4}
\Gamma^*(r) = \frac{3}{4 \pi r^3} \int_{0}^{r} 4 \pi r'^2 \Gamma(r') dr'
\ee
This function produce an artificial smoothing of 
$\Gamma(r)$ function, 
but it correctly 
reproduces global properties (CP92).

As we said, $\Gamma(r)$ (and $\Gamma^{*}(r)$) is the 
average density computed in spherical 
shell.  
In such a way we eliminate from the statistics
the points for which a sphere of radius {\it r} is not 
fully included within the sample boundaries. 
This prescription allows us to 
avoid any weighting scheme, i.e. 
any assumption in the treatment of
the boundaries conditions.
Of course in 
doing this, we have a smaller number of points 
and we stop our analysis  at a  smaller depth than that
of other authors (
\cite {bah83},
\cite {po92}, 
\cite {pli92},
\cite {ca92})
In fact, we have to limit 
our analysis  
to an {\it effective} depth 
$R_{s}$ that is of the order of the radius of the maximum
sphere fully contained in the sample volume 
(CP92; 
see also \cite {sym97}).
We have studied the behavior of $\:\Gamma(r)$ and $\:\Gamma^*(r)$
in the samples of {\it Table 1}. 
\begin{table}
\centering
\begin{tabular}{|c|c|c|c|c|c|c|}\hline
     &  &    &         &       &      &    \\
\rm{Sample} & N & $\Omega$ & $<n> $ & $d_{lim}$  & $R_{s}$&$r_{0}$ \\
            &   & $sr$ & $h^{3}Mpc^{-3}$& $(h^{-1}Mpc)$&$(h^{-1}Mpc)$&
$(h^{-1}Mpc)$ \\
\hline
\hline
BS83 sub north & $53$  & $3.1$  & $4.2 \cdot 10^{-6}$ 
&$230$ &$70$ &$27\pm 2$\\
     &   &  &   &      &   &            \\
\hline
     &  &   &   &      &   &             \\
Postman s.s. north & $120$  & $3.1 $  & $ 9.5 \cdot 10^{-5}$ 
&$230$&$70$& $26 \pm 2$\\
     &   &  &   &     &           &       \\
\hline
     &   &  &    &     &            &       \\
Postman s.s.south & $88$  & $1.7 $  & $ 1.3 \cdot 10^{-5}$ 
& $ 230$&$53$& $17\pm 2$  \\
     &   &  &   &      &         &       \\
\hline
     &  &   &   &     &            &        \\
ACO sub & $91$    & $2.2$  & $ 3.7 \cdot 10^{-5}$ 
&$150$&$50$&$17\pm1$\\
\hline
\end{tabular}
\caption{features of the various samples analyzed;
column I: the catalog from which the sample has been extracted;
column II: number of clusters in the sample;
column III: solid angle of the sample;
column IV: density of the sample;
column V: sample extension;
column VI: radius of the maximum sphere fully included;
column VII: the correlation length}
\label{pert}
\end{table}
The results 
are shown in Fig.2.
In Fig2a we have reported $\Gamma^{*}(r)$ 
for the north and 
south galactic part of Postman statistical subsample.
\bef
\epsfysize 10cm
\centerline{\epsfbox{fig2n.post}}
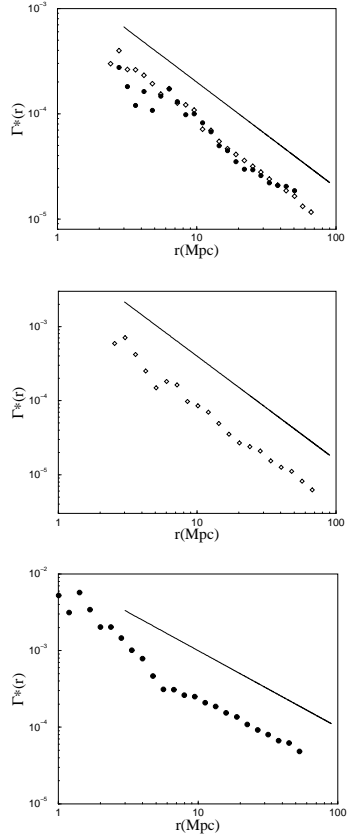
\caption{The behaviour of  the conditional density $\Gamma(r)$ 
and average conditional density 
$\Gamma^{*}(r)$ for 
a) Abell catalog ,Postman sample 
respectively north (diamonds) and south galactic parts (filled 
circles)
b) Abell catalog ,BS83 sample north galactic part (diamonds), 
c) ACO catalog sample (filled circles).
$\Gamma(r)$, $\Gamma^{*}(r)$ are power law up to the radius 
of the greatest sphere fully included in the sample.
The reference line has slope $\gamma \approx 1$  }
\eef

The south galactic part has a smaller solid angle 
with respect the north one and by consequence 
a smaller $R_{s}$. 
A well defined power law behavior is detected up
to the sample limit without any tendency towards homogenization.
The codimension is, with good accuracy 
\be
\label{g5}
\gamma = 3-D \approx 1.0 \pm 0.1
\ee
so that $\:D \approx 2.0 \pm 0.1$ up 
$\approx 70 h^{-1} Mpc$ for 
the north part and up to $\approx 50 h^{-1} Mpc$ for the south.
The result is in 
good 
agreement with those 
from BS83 sample (Fig.2b).
The north galactic 
part is fractal with dimension 
$D\approx 1.7 \pm 0.2$ 
up to $r\approx 70 h^{-1} Mpc$.
We have not reported the analysis 
for the BS83 south galactic part, because 
it gives a very noisy result, due to the 
very poor statistics of the sample.
The lower statistics of the BS83 sample 
with respect Postman one is also the 
reason for the little difference 
in the estimation of the fractal dimension 
$D$ of the two samples.
We note that the codimension 
$\gamma$ for these samples 
has a lower value with respect the 
standard determination; 
we refer the reader 
to next section for a explanation 
of this effect.
In Fig.2c we show 
the results from ACO sample.
The $\Gamma(r)$ is a power law with exponent
$D \sim  2.0 \pm 0.1$ up to $\approx 50 h^{-1} Mpc$.
The fluctuations at $r < 10 h^{-1} Mpc$
are due to the fact that these distances 
are of the order of the minimum average distance 
between clusters ($\approx 15 \hmp$ for 
Postman and BS83 and 
$\approx 8 \hmp$ for ACO sample), 
so that we can then interpret them 
as a effect of poor statistic.
In the correct regime, 
at distances $>10 h^{-1} Mpc$, where the samples
are statistically
significative, the slope of $\Gamma(r)$ is almost the same for the
Postman sample and for the ACO and, with a slight 
difference because the poor statistics, for the BS83 sample.
All the samples investigated have, then, 
consistent statistical
properties, i.e. they show a clear 
behaviour of the correlation function. 
A power law behaviour corresponds to 
long range correlations, which cannot 
be produced by random incompleteness 
in the sample, if they are not present.
Long range correlations ({\it fractal}) 
can be only destroyed by incompleteness, 
but not produced by it.

Hence the samples show well defined statistical properties, i.e.
they are {\it statistically fair} samples;
the results of the analysis is that 
they are not homogeneous samples, but, on the contrary, fractal.

\section{$\:\xi(r)$ analysis}

CP92 clarify some crucial points of the
standard CF analysis, and in particular they discuss the meaning
of the so-called {\it "correlation length"}
  $\:r_{0}$
found with the standard
approach (
\cite{da83})
and defined by the relation:
\be
\label{x1}
\xi(r_{0})= 1
\ee
where
\be
\label{x2}
\xi(r) = \frac{<n(\vec{r_{i}})n(\vec{r_{i}}+ \vec{r})>_{i}}{<n>^{2}}-1
\ee
is the two point correlation function used in the standard analysis.
As we said, if the average density is not a well defined intrinsic 
property of the system, the analysis with
$\xi(r)$ gives spurious results.
In particular, if the conditional density
$\Gamma(r)$ (or $\Gamma^{*}(r)$) is a power law, the system is 
{\it fractal} and the average density is simply related 
to the sample size.
In this case, if one derives a correlation length 
$r_{0}$, comparing 
the average density-density correlation 
to the square of average density, 
one obtains 
simply a fraction of the effective sample radius $R_{s}$.
In other words, if a system has a correlation
function $\Gamma(r)$ that is a power law, 
the system is self-similar and it 
has no {\it reference values} 
(like the average density) with
respect to which one can define what is big or small.
Hence, computing a correlation 
length with respect the average 
density is meaningless.
As pointed out in Baryshev et al. 1994
\cite {ba94}, 
this is true also for a system that has 
a characteristic length, as for 
example a system with correlation function 
$\Gamma(r) = \Gamma_{0} e ^{-r/r_{0}}$.
In this case, the characteristic length $r_{0}$ 
is related to the behaviour of the function and 
not the prefactor $\Gamma_{0}$; 
the characteristic length is not defined by 
the condition $\Gamma(r_{0}) = 1$.

 Following CP92, the expression of the $\:\xi(r)$ in the case of
fractal distribution, is:
\be
\label{x3}
\xi(r) = ((3-\gamma)/3)(r/R_{s})^{-\gamma} -1
\ee
where $\:R_{s}$ (the effective sample 
radius) is the depth of 
the spherical volume where one computes the
average density from Eq. (\ref{l2}).
From Eq. (14) it follows that

i.) the so-called correlation
length $\:r_{0}$ (defined as $\:\xi(r_{0}) = 1$)
is a linear function of the sample size $\:R_{s}$
\be
\label{x4}
r_{0} = ((3-\gamma)/6)^{\frac{1}{\gamma}}R_{s}
\ee
and hence it is a quantity without any correlation meaning but it is
simply related to the sample size.

ii.) the amplitude of the $\xi(r)$ is:
\be 
\label  {x5}
A(R_{s}) = ((3-\gamma)/3)R_{s}^{\gamma} 
\ee

iii.) $\:\xi(r)$ is a power law only for 
\be
\label{x6}
((3-\gamma)/3)(r/R_{s})^{-\gamma}  >> 1
\ee
hence for $\: r \ltapprox r_{0}$: for larger distances
there is a clear deviation from a
power law behavior due to the definition of $\:\xi(r)$.
However usually 
$\xi(r)$ is fitted with a power law in the 
range $ 0.5 r_{0} \ltapprox  
r \ltapprox 2 r_{0}$.
The result is one obtains
 a greater exponent than 
$ 3 - D $. This is the reason why 
the usual estimation of the 
exponent $\gamma$ 
of $\xi(r)$ is $\approx 1.7$, different 
from $\gamma \approx 1$, corresponding
to $D \approx 2$, that we found by mean of  
the $\Gamma(r)$ analysis.
Moreover we point out that there is another spurious 
effect in the calculation of the $\xi(r)$.
The $\xi(r)$ is estimated by means of the expression
\be
\label{x7}
\xi(r) = \frac{N_{dd}(r)}{N_{pp}(r)} - 1
\ee
where $N_{dd}(r)$ is the number of
data pairs at separation $r$
and $N_{pp}(r)$ is the number of pairs 
at separation $r$ for a random distribution 
with the same density and geometry of the survey.
Eq.(18)
corresponds to compute the density-
density correlation not only in spheres
fully contained in the sample volume, but also in 
portions of sphere.
This produce an artificial homogenization,
for distances $r$ greater than the 
radius of the greatest sphere fully 
contained in the sample.
For these separations, in fact, the 
density -density correlation is, of course,  
computed only in portions of sphere and 
this corresponds to assume that the density, 
found in the portion, is the same in the all  
solid angle. 
To avoid this effect, we have computed the $\xi(r)$ 
only up to $R_{s}$. 
We have studied the $\:\xi(r)$ in our samples
of clusters. In fig. 3 
\bef
\epsfysize 10cm
\centerline{\epsfbox{figxt.post}}
\caption{
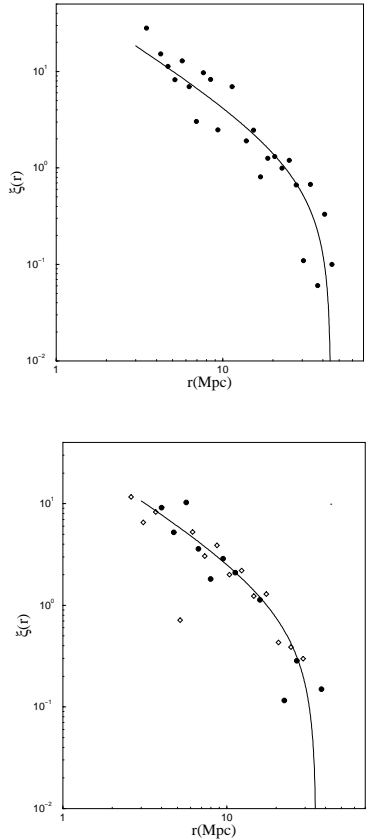
The $\xi(r)$ function for the
a) Postman sample (north) (filled circles)
b) Postman sample (south) (filled circles) and ACO (diamonds).
The reference line is the functional form of eq.(14) with
$\gamma \approx 1$}
\eef
we have reported the 
$\xi(r)$ for the Postman sample (north and south) 
and for the ACO sample.
We have fitted the experimental points 
with the functional form of eq.(\ref{x3}).
 We found that 
$\:\gamma \approx 1$ for the all samples 
analyzed. The corresponding $r_{0}$ are 
reported in the {\it Table 1}; for the 
Postman sample north we found 
$r_{0}\approx 26 \pm 2$ and for BS83 sample 
$r_{0} \approx 27\pm 2$
These two samples have the same $r_{0}$ as we 
expect, noting that they have the same $R_{s}$.
The values found are in agreement with eq.(15).
The same conclusions hold for the other
two samples: Postman south and ACO.
Both of them have $R_{s}$ ($\approx 50 h^{-1} Mpc$) and  
the same $r_{0}$ ($\approx 17 h^{-1} Mpc$), according to eq.(15). 
In conclusion $r_{0}$ is  
simply a fraction of the sample size $R_{s}$, 
without any physical meaning for the 
properties of the samples.

\section {Conclusion}

We have analyzed the properties of the various Abell/ACO 
cluster samples. The statistical analysis, performed without 
any a priori assumptions, shows that Abell samples 
are fractal 
with fractal dimension 
$D \approx 2 $ for Postman sample and 
for BS83 sample up to $\approx 70 h^{-1} Mpc$
 and for ACO sample 
up to $\approx 50 h^{-1} Mpc$.
The different limiting distance of the analysis 
in the various samples corresponds to the 
radius of the greatest sphere fully included in the sample, 
that is different in the various samples. 
This limitation allows us to avoid 
any  weighting schemes or assumptions in the 
analysis.
The result of the statistical analysis 
 is that no tendency towards homogenization is 
detected within the sample limits.
The so-called correlation length $r_{0}$ 
derived from the $\xi(r)$ analysis, 
is simply directly proportional to the sample size $R_{s}$ 
and then, it is meaningless
with respect the correlation properties of the system.

 This result is in agreement with the 
correlation analysis, performed measuring the 
conditional average density, $\Gamma(r)$, that is a 
power law (${\it fractal}$) within the sample limits.
Moreover our results on cluster samples are in remarkable agreement 
with the same analysis performed on various  galaxy redshift surveys
(CfA1, Perseus Pisces, LEDA, ESP)
as one expects for a fractal distribution of galaxy and clusters.
The mismatch between galaxy and cluster correlation 
is just due to the mathematical inconsistency of the 
use of $\xi(r)$ and the correct analysis, in terms 
of $\Gamma(r)$ and $\Gamma^{*}(r)$, shows that 
correlations of clusters are just
the continuation at larger scales of galaxy 
correlations.
We conclude that galaxies and clusters are two different 
representations of the same self similar structure.

Galaxy clusters extend the correlations of galaxies to 
deeper depth. 
In this respect cluster distribution 
represents a coarse grained 
representation of galaxies; 
it is the same self-similar distribution, 
but sampled with a large scale resolution.
i.e. considering 
a cluster of galaxies as 
a single object, without distinguish the 
structure in it.
Hence, we can study the clusters distribution simply 
performing 
a coarse graining on galaxy distribution. 
Usually clusters are identified with 
some criteria, that are different according to 
different observers.
On the contrary, in this way, 
we can make the analysis independently 
from the definition of {\it cluster}, {\it supercluster}
etc., but one has 
to know the complete galaxy 
survey over which performing 
the coarse graining procedure. 
The same considerations hold, of course, for the 
void distribution: the void distribution is, 
in fact, just 
the complement of the matter one. 
In conclusion, our 
methods have the advantage to be 
independent from the 
nature of the distribution considered, and 
they gives a quantitative way to detect self 
similar properties, whether they exist, at different 
scales.
\newpage
{\bf Acknowledgements}

We thank prof. L. Pietronero for 
useful discussions and 
enlightening suggestions.
We are also grateful to A.Amendola,
A.Gabrielli, H. Di Nella, 
H.Andernach and R. Cafiero; to R. Capuzzo Dolcetta 
for useful comments and collaboration.
This work has been partially supported by 
the Italian Space Agency (ASI).





\end{document}